Comparison of data on Mutation Frequencies of Mice Caused by Radiation

- Low Dose Model -

Yuichiro Manabe,

Masako Bando[1,2]

Division of Sustainable Energy and Environmental Engineering, Graduate School of Engineering, Osaka University, 2-1 Yamada-oka, Suita, Osaka 565-0871, Japan

[1]Jein Institute of Fundamental Science, Venture Business Laboratory, Kyoto University, Yoshida, Honmachi, Sakyo-ku, Kyoto-shi, 606-8501, Japan

[2]Institute of Fundamental Physics, Kyoto University, Kitashirakawa oiwake-cho, Sakyo-ku, Kyoto-shi, 606-8502, Japan

Abstract

We propose   LD(Low Dose) model, the extension of LDM model which was proposed in the previous paper [Y. Manabe et al.: J. Phys. Soc. Jpn. 81 (2012) 104004] to estimate biological damage caused by irradiation. LD model takes account of all the considerable effects including cell death effect as well as proliferation, apoptosis, repair. As a typical example of estimation,   we apply LD model to the experiment of mutation frequency on the responses induced by the exposure to low levels of ionizing radiation. The most famous and extensive experiments are those summarized by Russell and Kelly [Russell, W. L. & Kelly, E. M: Proc. Natl Acad. Sci. USA 79 (1982) 539-541], which are known as "Mega-mouse project". This provides us with important information of the frequencies of transmitted specific-locus mutations induced in mouse spermatogonia stem-cells. It is found that the numerical results of the mutation frequency of mice are in reasonable agreement with the experimental data: the LD model reproduces the total dose and dose rate dependence of data reasonably. In order to see such dose-rate dependence more explicitly, we introduce the dose-rate effectiveness factor (DREF). This represents a sort of preventable effects such as repair, apoptosis and death of broken cells, which are to be competitive with proliferation effect of broken cells induced by irradiation.

## 1. Introduction

Nowadays, it is one of the most controversial issues how low-level radiation hurts biological objects. If it is merely a physical process, the frequency of radiation-induced mutations is proportional to total dose irradiation because in most cases biological damage begins with the mutation of living cells which is caused by ionization. Then a most reasonable hypothesis may be based on physical procedure of "a stimulus and response". Indeed in 1927, Herman J. Muller studied the effect of X-rays on Drosophila[1], and showed, together with the works followed, a linear dependence on total radiation dose exposure to the number of mutations frequency, without any threshold effects. This observation caused a strong impact not only on the scientific community but on whole society and has led to official adoption of what is called "LNT hypothesis" for approximately a decade, until W. Russell, of the Oak Ridge National Laboratory, proposed to test its validity in mice. The results obtained by Russell did not confirm the LNT; the lower the level of dose rate, the more the time needed to produce the same effect. This indicates the existence of mechanisms by which cells can be protected against irradiation.

However this is not enough to clarify the situation and there arises many arguments in between two extremes; there have been arguments to support lower (hormesis effects, for example) or higher (coming from bystander effects) risk. This is because the subsequent biological processes make sometimes preventable effects on the one hand, and on the other hand yield enhancement of mutation of living cells. In any case there are various kinds of mechanism operating on living objects, and it is necessary to construct a model to make systematic analysis taking account of every possible effect in order to make quantitative estimation.

In a separate paper[2] we propose a mathematical model to estimate biological damage caused by radiation, which we call LDM (Low Dose Meeting) Model.

However the original LDM model is not applicable to the case of more general situation, although it shows how the recovery effects suppress the rapid proliferation of broken cells in realistic living objects. For example, in applying LDM model to realistic processes, we have to take account of additional effect, namely cell death effect caused by irradiation.

The aim of this paper is to reformulate LDM model by taking account of all conceivable biological effects under more general situations in living objects. Hereafter we call this simply "LD (Low Dose) model". Then we apply this LD model to the realistic process by taking an example of mutation frequency of mice. The results can be compared with the experimental data. We here focus on the most famous experiments known as "Mega-mouse project" [3,4].

In section 2, we propose general formalism of LD model as an extended version of LDM model by taking account of such additional effect as cell death, cell density dependence and so on. In section 3, we summarize the experimental data which we are going to analyze. Section 4, we make approximation of general LD model to apply the experimental data. The numerical results are made in section 5, and we further concentrate our attention on the dose rate effectiveness factor (DREF) which represents the dose-rate dependence of mutation frequency in section 6. Section 7 is devoted to summarized discussions.

2. Formalism of LD model

Consider a system of cells, a tissue or an organ (hereafter we call it symbolically "tissue") with its capacity $\left(N_n\right)_{\max} = K$, the maximum number of cells of this system. The diagram of Fig.1 shows the dynamical situation of stimulus-response of the cells caused by irradiation (straight wave lines) together with other stimulus (dotted wave lines). In living object there exist various types of functions, proliferation, repair and apoptosis effects. Further we take account of cell death effects caused by irradiation as well.

Living objects are receiving a certain sort of stimulus from their surroundings just after they are born. Their effects may be balanced with their preventable effects such as repair and apoptosis, and for long time during those period the number of broken cells, $N_b$ tends to a certain number maintaining stationary state until artificial irradiation starts. Let us consider such realistic situation and take account of all the effects which we have to consider.

Suppose that a tissue contains normal cells with its number $N_n$ and broken cells, $N_b$ and at $t = 0$ it is exposed by radiation with a constant dose-rate $r$. The risk estimate may be related to the number of broken cells which may turn to a cancer tumor.

The cell numbers, normal and broken ones, $N_n$ and $N_b$, respectively, obey the following equations;

$$\frac{d}{dt}N_n(t) = h(N_n)\left[\alpha_n N_n(t) + \mu_r N_b(t)\right] - g(N_n)c(r(t) + r_{eff}) - g(N_n)d_n \cdot (r(t) + r_{eff}),$$

$$\frac{d}{dt}N_b(t) = g(N_n)c \cdot \left(r(t) + r_{eff}\right) - g(N_b)d_b \cdot \left(r(t) + r_{eff}\right) + \alpha_b N_b(t) - \left(\mu_r + \mu_a\right)N_b(t)$$

$$= g(N_n)c \cdot \left(r(t) + r_{eff}\right) - g(N_b)d_b \cdot \left(r(t) + r_{eff}\right) + \mu N_b(t),$$

$$h(N_n(t)) \equiv 1 - \frac{N_n(t)}{K}, g(N_n(t)) \equiv 1 - \frac{1}{1 + \beta N_n(t)}, g(N_b(t)) \equiv 1 - \frac{1}{1 + \beta N_b(t)}.$$

$$\mu \equiv \alpha_b - \mu_r - \mu_a \quad (2.1)$$

The notations $\alpha_n, \alpha_b, \mu_r$ and $\mu_a$, are proliferation rate of normal, broken cells, the rates of inducing repair and apoptosis of broken cells, respectively.

We here introduce the suppression factor $h(N_n(t)) = \left(1 - \dfrac{N_n(t)}{K}\right)$ to control the number of normal cells. The characteristic feature of normal cell is that they arrest their proliferation when the number of normal cells approaches its maximum $K$, namely $h(N_n(t)) \to 0$ when $N_n(t) \to K$. On the other hand, broken cells never stop their proliferation.

The function $g(N_n)$ depends on the number of normal cells in general. However, let us remind that we here use the unit Gy for the strength of irradiation. This represents the unit of absorbed dose, the absorption of one joule of energy, inducing ionizing radiation, per kilogram of matter. Thus in terms of Gy, the irradiation strength rate $r(t)$ deposits a corresponding increment of energy per time in unit volume of tissue, and thus the derivative of total number of broken cells, is proportional to the amount of irradiation strength rate $r(t)$ only, so far as we consider the case where cell density is high enough. Such kind of treatment may be similar to the situation in which nuclear physicists often employs the concept of nuclear matter which is defined as an idealized system consisting of a huge number of protons and neutrons with finite density. Thus $g(N_n)$ is independent of cell density unless normal cell number is extremely small. On the contrary if the cell density becomes very small, namely the number of normal cells is very small and the radioactive energy deposit is not fully poured into the breakdown of normal cells the ratio of which is in proportion to the number of normal cells. So $g(N_n)$ should show the following asymptotic behavior,

$$g(N_n) \xrightarrow[\beta N_n \ll 1]{} \beta N_n,$$
$$\xrightarrow[\beta N_n \gg 1]{} 1. \quad (2.2)$$

As such examples we can take

$$g(N_n) = 1 - \frac{1}{1 + \beta N_n},$$
$$g(N_n) = 1 - \exp(-\beta N_n). \quad (2.3)$$

Here $\beta^{-1}$ represents the critical number which divide the region where $g(N) \sim 1$ and $g(N) \sim N$. The breaking coefficient $c$ in Eq. (2.1), which has been added in back of $g(N_n)$, depends on the kinds of the living object, organ or tissues under consideration. It is a breaking coefficient to the irradiation strength rate $r(t)$. This is what we call radiosensitivity. In general, the coefficient $c$ might be determined by radiation cross section of cells, cell density, and the related surrounding conditions.

$d_n$ and $d_b$ are the rates of death of normal and broken cells caused by artificial irradiation $r(t)$, respectively. For the case where an artificial irradiation starts at t=0,

$$r(t) = \begin{cases} 0 \, (t < 0) \\ r(t \geq 0) \end{cases}.$$

We also multiply $g(N_n) = 1 - \dfrac{1}{1 + \beta N_n(t)}$ by the coefficient $d_n$, $g(N_b) = 1 - \dfrac{1}{1 + \beta N_b(t)}$ by $d_b$.

We introduce $r_{eff}$, effective dose rate, to which we convert the stimulus coming from those other than artificial irradiation. It can be taken as almost constant since it comes from the environment where living object lives. Note that the stimulus from other sources, physical, chemical or biological ones break normal cells into broken ones from the very starting point of living objects (at birth of living object). The terms coming from repair and apoptosis effects, which we can totally call "preventable effect"[5], however, usually overwhelms proliferation effect, unless living objects would die.

## 3. Experimental Data

We here focus our attention on the accumulated data of mutation frequencies caused by irradiation in male mice, which includes those at the lowest dose rate so far tested. Their data are summarized by Russell and Kelly[4]. Their most important finding is that radiation induced mutation frequency is not only determined by the total dose but also the very dose rate, namely at low dose rates the risk is relatively less than that the case of high dose rates. According to ref. 4, the data reported there were obtained by the so-called "mouse specific-locus test (SLT)"[6]. This uses seven visible markers and permits the detection of mutations involving any of the seven gene loci in the first-generation offspring of the irradiated parent.

Table 1

The results from treated and control adult males are to be compared in order to see the pure effects coming from artificial irradiation. The radiation was $\gamma$-rays from a $^{137}$Cs or $^{60}$Co sources or X-rays. In addition, the experimental results which had been done by several authors were summarized in Table 1, where we label those data by A to O. We here express the values of the data in terms of the scale unit recently used. In The original figure of ref. 4 is shown as the data sets of the total dose dependence of mutation frequency (Fig.2), which indicates that the number of mutations is not determined merely by total dose. Actually they divided the data into two groups, acute and chronic irradiation cases (indicated by the category in Table 1), where (M, N, O) belong to acute group and the rest (A to L), to chronic group.

Fig. 2

The figure (Fig.2) of ref.4 indicates only 11 data points; they reduced to 11 out of 15, namely 6 points, E and L with total dose 6 Gy and A, D, F, I with total dose 3 Gy, are combined into 2 groups and the data points are indicated 2 by taking their averages. However the combined data points are composed of different dose-rate data and it is not be a proper way to take their average. Later we shall discuss on this point.

It would be convenient to take the unit of mutation frequency $F$ and total dose $R$ expressed in terms of $10^{-5}$ and Gy units, respectively. We shall use those units in section 4.

Now the finding of Russell and Kelly is that the slope for acute irradiation is much steeper than the one of chronic irradiation: the ratio of the chronic irradiation is almost 30% of that for acute irradiation. They gave the best estimate of the ratio of number of mutation for chronic and acute irradiation with straight lines,

$$F_{chronic}(R) = F_{control} + \alpha_c R,$$
$$F_{acute}(R) = F_{control} + \alpha_a R, \qquad (3.1)$$

with the intercept $F_{control} \doteqdot 0.81 \times 10^{-5}$, $\alpha_c \doteqdot 0.73 \times 10^{-5}$ [1/Gy], $\alpha_a \doteqdot 2.9 \times 10^{-5}$ [1/Gy] .

Those straight lines of best fit are also shown in Fig. 1. We clearly see the indication of dose-rate effectiveness factor (DREF) which uses in some sense similar concept to dose and dose effectiveness factor (DDREF). Note that, however, some data points still deviate from the above two lines. We shall see the situation more clearly in section 4.

### 4. LD Model with high density approximation

The data summarized by Russell and Kelly[4] which we are going to analyze indicate that $\dfrac{N_b}{N_n}$ is of the order of $10^{-5}$. Therefore we can take the following form

$$g(N_n(t)) = 1 - \frac{1}{1 + \beta N_n(t)} \sim 1, \, g(N_b(t)) = 1 - \frac{1}{1 + \beta N_b(t)} \sim \beta N_b(t)$$

as $\beta N_n(t) \gg 1$ and $\beta N_b(t) \ll 1$. Therefore we can approximate Eq. (2.1) as

$$\frac{d}{dt} N_n(t) = \left(1 - \frac{N_n(t)}{K}\right)\left(\alpha_n N_n(t) + \mu_r N_b(t)\right) - c\left(r(t) + r_{eff}\right) - \tilde{d}_n\left(r(t) + r_{eff}\right),$$

$$\frac{d}{dt} N_b(t) = c\left(r(t) + r_{eff}\right) + \left(\alpha_b - \mu_r - \mu_a\right) N_b(t) - \tilde{d}_b\left(r(t) + r_{eff}\right) N_b(t)$$

$$= c\left(r(t) + r_{eff}\right) - \left\{\mu + \tilde{d}_b\left(r(t) + r_{eff}\right)\right\} N_b(t), \qquad (4.1)$$

$$\mu \equiv \mu_r + \mu_b - \alpha_b, \, \tilde{d}_n = d_n\beta, \, \tilde{d}_b = d_b\beta.$$

Now we apply LD model to estimate mutation frequency $F(t)$. In this case the system is specific-locus which consists of $N_n$ normal cells. And some of normal cells are broken by irradiation changing to $N_b$ broken cells. The frequency of transmitted specific-locus mutations induced by irradiation $r$ in mouse supermatogonial stem cells, namely $F(t)$ is defined as $\dfrac{N_b}{N_n}$. As seen in Table. 1, $F(t)$ is of order of $10^{-5}$. In addition

$$N_n(t) \sim N_n(t=0) - \Delta N_n(t).$$

The second term comes from the loss of normal cells due to the change into broken cells. Looking at the situation of experimental conditions made by Russell and others, the tissue under consideration enough grow up as adult and $N_n$ almost approaches to its maximum $K$. In the second term the one coming from the breakdown is of $10^{-5} \cdot N_n$ judging from the mutation rate so long as the irradiation rate is small enough to the keep the death term. This can be reasonable since the mice are alive to become adult and can make offsprings after the irradiation is stopped. Thus we can assume that $N_n$ is kept almost constant even if some of them are lost.

We compare the predicted values with the data summarized in Table 1. Since $N_n(t) >> N_b(t)$, Eq. (4.1) can be truncated into the following form in terms of the mutation frequency, $F(t)$

$$\frac{d}{dt}\left(\frac{N_b}{N_n}\right) = \frac{c}{N_n}\left(r(t)+r_{eff}\right) - \left\{\mu + \tilde{d}_b\left(r(t)+r_{eff}\right)\right\}\left(\frac{N_b}{N_n}\right).$$

$$\Leftrightarrow \frac{d}{dt}F(t) = C\left(r(t)+r_{eff}\right) - \left\{\mu + \tilde{d}_b\left(r(t)+r_{eff}\right)\right\}F(t),$$

$$C \equiv \frac{c}{N_n}. \tag{4.2}$$

The solution of Eq. (4.2) for $t \geq 0$ is

$$F(t) = -\left(F(0) - \frac{C\left(r + r_{eff}\right)}{\mu + \tilde{d}_b\left(r + r_{eff}\right)}\right)\left(1 - \exp\left(-\left(\mu + \tilde{d}_b\left(r + r_{eff}\right)\right)t\right)\right) + F(0),$$

$$R\left(t\right) \equiv rt. \ (4.3)$$

For $r = 0$, where there is no artificial irradiation, the solution of equation is written as,

$$F(t) = \frac{Cr_{eff}}{\mu + \tilde{d}_b r_{eff}}\left(1 - \exp\left(-\left(\mu + \tilde{d}_b r_{eff}\right)(t - t_0)\right)\right), (4.4)$$

with the boundary condition $F\left(t = t_0\right) = 0$ (the time $t_0$ corresponds to the time when a living object was born. ).

An example of behavior of Eq. (4.4) is seen in Fig. 3. We see that $F\left(t\right)$ in Eq. (4.4) tends to some constant, whose value is easily obtained from Eq. (4.4),

$$F(t) \xrightarrow[t_0 \to -\infty]{} \frac{Cr_{eff}}{\mu + \tilde{d}_b r_{eff}}. (4.5)$$

In actual situation we set $t = 0$ at the time when irradiation starts. Thus Eq. (4.5) corresponds to $F(t = 0)$ of Eq. (4.3), namely

$$F\left(0\right) = \frac{Cr_{eff}}{\mu + \tilde{d}_b r_{eff}}. (4.6)$$

$F\left(0\right)$ can be interpreted as the mutation frequency of control mouse of Tab.1.

Fig. 3

Now we can rewrite Eq. (4.3),

$$F(t) = -\left( F(0) - \frac{C\left(r_{eff} + r\right)}{\mu + \tilde{d}_b\left(r_{eff} + r\right)} \right)\left\{ 1 - \exp\left( -\left(\mu + \tilde{d}_b\left(r_{eff} + r\right)\right)t \right) \right\} + F(0),$$

$$= \frac{c'r}{\mu'}\left(1 - \exp(-\mu't)\right) + F(0),$$

$$c' \equiv C - F(0)\tilde{d}_b, \; \mu' \equiv \tilde{d}_b r + \frac{C}{c'}\mu. \; (4.7)$$

In order to make numerical calculation we take each unit as follows;

$[F(t)] = 10^{-5}$ : Mutation frequency in $10^{-5}$ per locus,

$[R] \equiv \mathrm{Gy}$ : total dose,

$[r] \equiv \mathrm{Gy/hr}$ : dose rate per hour,

$[c] \equiv 1/\mathrm{Gy}$ : effective mutation rate per Gy and hour in unit $10^5$,

$[\mu] \equiv 1/\mathrm{hr}$ : effective multiple rate per hour,

$\mu^{-1}$ : relaxation time in hour   and   $\tau = \mu t$ : time index in terms of unit of relaxation time $\mu^{-1}$.

The terms coming from repair and apoptosis effects, which we can totally call "preventable effect"[5], however, usually overwhelms proliferation effect, unless living objects would die. Indeed the data summarized by Russell and Kelly[4] which we are going to analyze, we will see that the index number $\mu$ in Eq. (2.1) is found to be positive.

## 5. Numerical calculation and results

Let us make numerical calculation of the time dependence of mutation frequency $F(t)$ of Eq. (4.7). We have three parameters $\mu$, $\tilde{d}_b$ and $C$ to be determined from the experimental data. The number of data sets is 15, where we select 9 data sets among 15 of Table 1, two of which are the ones denoted by E and L and 4 of which correspond to A, D, F, I with total dose 6 Gy and 3 Gy, respectively. These two groups were treated as single-data point in Fig.1 in ref.4. The LDM model II indicates that the dose rate dependence is very important. Unfortunately the dose rates of the groups are different one another. In order to compare the prediction of LDM with each data we need error bars for each data, which are not listed in Table 1 and we have to read off them from Fig.2. As for the group data (A, D, F, I) and (E, L), it is impossible to extract each error bars. Without any error bars, the degree of reliability would be lower, so we first select 9 data sets, B, C, J, H, G, K, M, N, and O. Later we shall show the comparison of our model with all the date including those without any error bars.

The three parameters $\mu$, $\tilde{d}_b$ and $C$ are determined by using the chi-squared fit procedure so as to match the estimation for the observed data, and as a result we get,

$$\mu = 3.13 \times 10^{-3} \ [1/\text{hr}],$$
$$\tilde{d}_b = 1.00 \times 10^{-1} \ [1/\text{Gy}],$$
$$C = 2.91 \times 10^{-5} \ [1/\text{Gy}], \ (5.1)$$

with the control data $F(t=0)=1$.

Now, let us compare 9 observed data sets with the estimated results using the above three fixed parameters, $\mu$, $\tilde{d}_b$ and $C$. Since the behavior of mutation frequency $F$ depends on dose rate explicitly, we divide the whole data points into groups with the same order of dose rate, acute ($10^{+1}$ Gy/h), chronic with low ($10^{-3}$ Gy/h ) and lowest ($10^{-4}$ Gy/h) dose rate, respectively. The calculated results are shown in Fig.4-a, b and c for acute, intermediate and chronic dose rate groups, respectively. Note that the predicted curves are indicated the total dose dependence for corresponding dose rates.

Fig.4

It turns out that all the predicted values are within their error bars. Note that we have here assumed that the observation (or more exactly the first-generation of offspring is created by mating) was done just after the continuous constant irradiation stopped, namely the number of offspring is fixed just after irradiation stops. The total time $T$ of Fig.4 is therefore estimated as $T = \text{(total dose)} / \text{(dose rate)}$ from total doses and dose rates of Table 1. It is seen that the predicted value of mutation frequency estimated from LD model depends strongly on dose rate, and the individual data points in the figure correspond to the mutation frequency after the time interval T, during which the total amount of dose becomes the corresponding value of each data point. It turns out that our LD model reproduces whole structure quite well, all the predicted values are just within error bars so far as for the 9 data with definite dose rates groups, namely B, C, J, H, G, K, M, N, and O. For the other 6 data, denoted by E and L with total dose 6 Gy and A, D, F, I with total dose 3 Gy, they are controversial because they are combined into groups classified only by their total dose, though they correspond to different dose rates. In the next section we shall comment on those data points and discuss more about the dose rate dependence.

### 6. More about DRFF

Now that the mutation frequency estimated from LDM model II depends strongly on dose rate, we display individual data points in a figure to see the whole scope of 9 data points with error bars indicated (Fig.5).

Fig. 5

In this figure the horizontal axis is taken logarithmic scale. Since the mutation frequency is dependent of dose rate, we further show the total dose dependence for 5 typical dose-rate $r$, $10^{+1}, 10^{-1}, 10^{+2}, 10^{-3}$ and $10^{-4}$ Gy/hr. This figure shows global picture of dose rate dependence of 9 data points. It turns out that our LD model reproduces whole structure quite well, which has been already confirmed in Fig. 4; all the data points are within the region of corresponding dose rates. Fig.4. also we can see that the asymptotic behavior of mutation frequency is

$$F(t) \xrightarrow{t \to \infty} \frac{C\left(r + r_{eff}\right)}{\mu + \tilde{d}_b\left(r + r_{eff}\right)}, (6.1)$$

so it is almost independent of the dose rate if dose-rate

$r$ is enough strong satisfying the following condition.

$$\frac{\mu}{\tilde{d}_b} << r + r_{eff} \Rightarrow F(t) \xrightarrow{t \to \infty} \frac{C}{\tilde{d}_b}. (6.2)$$

From the Eq. (5.1)

$$\frac{\mu}{\tilde{d}_b} = \frac{3.13 \times 10^{-3} \text{ [1/hr]}}{1.00 \times 10^{-1} \text{ [1/Gy]}} = 3.13 \times 10^{-2} \text{ [Gy/hr]} = 27.4 \text{ [Sv/year]},$$

$$\frac{C}{\tilde{d}_b} = \frac{2.91 \times 10^{-5} \text{ [1/Gy]}}{1.00 \times 10^{-1} \text{ [1/Gy]}} = 2.91 \times 10^{-4}, (6.3)$$

which is what we see in Fig. 5, if dose rate is strong enough the frequency depends on total dose only.

We have found that there is a certain critical value of dos-rate below which the mutation frequency $F(t)$ is not determined by total dose $R$ only but depends on the strength of irradiation $r$. For fixed $R$, dose-rate dependence of mutation frequency determines the DRFF, dose-rate effectiveness factor defined as,

$$DREF(r) = \frac{F\left(r >> 1[\text{Gy/hr}]\right)}{F(r)},$$

$$= \frac{\lim_{r \to \infty} \frac{c'r}{\mu'}\left(1 - \exp(-\mu'\frac{R}{r})\right) + F(0)}{\frac{c'r}{\mu'}\left(1 - \exp(-\mu'\frac{R}{r})\right) + F(0)}$$

$$\to \frac{\frac{c'}{d_b}\left(1 - \exp(-d_b R)\right) + F(0)}{\frac{c'r}{\mu'}\left(1 - \exp(-\mu'\frac{R}{r})\right) + F(0)}, (5.4)$$

$$c' \equiv C - F(0)\tilde{d}_b, \ \mu' \equiv d_b r + \frac{C}{c'}\mu.$$

In Fig. 6, we show the dose rate dependence of DREF. Solid, dashed and dotted lines correspond to the case of total dose $R = 10, 1, 0.5[\text{Gy}]$, respectively.

Fig. 6

Note that DDREF does not actually depend on dose rate itself. For more detailed discussion we would like to leave to another paper.

Now let us return to the data which we have so far excluded. If we include all the data points by adding six to nine ones as shown in Fig. 7, we see some discrepancy between experimental data and theoretical values, especially for the data points, E and L.

Fig. 7

To see the situation more clearly, we show in Fig. 8 the comparison between experimental data points with error bars and theoretical values of mutation frequency. Among them, four data, A, D, F, I may lie within the prediction range because error bar of combined data shown in Fig.2 is about 2 and each one of separated data points must be more than 2. However two data E and L, with completely different order of dose rates $10^{-4}$ and $10^{-1}$ [Gy/hr], is serious as they are far from theoretical data with deviations span being 4.4 and 8.5 respectively. If we compare the data points with $R = 6$ [Gy], E, L, N with dose rates $10^{-4}, 10^{-1}$ and $10^{+1}$ [Gy/hr], the experimental data of L is found to be the smallest. Such possibility is quite unlikely and is in contradiction to the statement made by Russell himself who pointed out that the lower the level of dose rate, the lower the mutation frequency becomes. To be more careful we cannot definitely conclude that LD model with the present parameters show best fit to reproduce the very low rate region $r \sim 10^{-4}$ Gy/h because the data with error bars, B and C, are a little far from the other three data A, D and E, and if we choose the latter data group more strictly and neglect the former, we may yet have room to take account more unknown effect to reproduce so-called hormesis phenomena. Indeed some authors insist that the mutation is minimum in the range of $60$ mGy/h $< r < 600$ mGy/h [11),12)]. This we would like to leave to our future task and to investigate more carefully.

Fig. 8

In any case the data for several $10^{-1}$ Gy/hr dose rate plays an important role and it is very critical to determine DRFF. There is only one data point indicated L, unfortunately. Experimental data for dose rate around several $10^{-1}$ Gy/hr may be highly desired to know the critical dose-rate which divides chronic and acute irradiation.

## 7. Conclusion and Discussion

We have shown that our mathematical model to estimate biological damage caused by radiation (LD Model) reproduces the well-known experimental data on mutation induction in mice reasonably. Accumulation of animal studies on the biological effects of radiation was extensively carried out after the World War II. Among them, the data summarized by Russell and Kelly[4] are well known as "Mega-mouse project", which are even now most frequently referred by many authors. We have here compared their data with our prediction obtained from our LD model. The calculated numerical results are found to be in reasonably agreement with the observed data. We clearly show that the mutation frequency does not depend merely on total dose but on dose rate as well. This indicates that we can estimate the dose-rate effectiveness factor (DREF) almost exactly if we get detailed information of the irradiation processes. It can open the window to estimate the radiation risk in a quantitative way; the LD model may be a first good example to be applicable for the estimation of radiation risks.

**Acknowledgements**

The authors would like to thank to K. Uno, K. Fujita, M. Abe, H.Utsumi and O. Niwa for encouragement and informing us many experimental data of frequency of mutation for various irradiation exposure cases. Also thanks are due to Ichikawa for his collaboration in the analysis in early stage and the members of LDM, for their various comments and discussions.

Table 1. Specific locus mutation rate (multiplied by $10^5$) data obtained in $A_s$ spermatogonia by use of the seven-locus test stock provided by ref. 4

Fig. 1. Stimulus response diagram of LD model.

Fig. 2. Mutation rates (multiplied by $10^5$) at seven specific loci in the mouse versus total dose. Original figure presented by ref. 4 is to be compared with the one with all the individual data points listed in Table 1 and the unit of dose is expressed in terms of roentgen unit. The straight lines are obtained by ref. 4, the best estimate of the ratio of mutation frequencies for chronic and acute irradiation.

Fig. 3. An example of time dependence of Eq. (4.2).

Fig. 4. Time dependence of mutation frequency dividing into each dose rate. The data points are indicated by those predicted at time T. T is determined by    which corresponds the time when it reaches the corresponding total dose of each data.

Fig. 5. Explicit dose rate dependence of mutation frequency $F$ can be easily seen by dividing the whole data points into groups with the same order dose rate groups, acute, chronic with $10^{-3}$ and $10^{-4}$ [Gy/hr] dose-rate groups, respectively. The calculated results are indicated by three lines, straight and dotted and dashed -dotted lines corresponding to acute ($5 \times 10$ [Gy/hr]), chronic with $5 \times 10^{-3}$ [Gy/hr]), chronic with ($5 \times 10^{-4}$ [Gy/hr]) groups, respectively. Nine data points are also indicated in the figure with error bars.

Fig. 6. Dose rate dependence of DREF. Solid line, dashaed line and dotted line

correspond to the case of total dose $R = 10, 1, 0.5 [\text{Gy}]$, respectively.

Fig. 7. Explicit dose rate dependence of mutation frequency $F$ can be easily seen by dividing the whole data points into groups with the same order dose rate groups, acute, chronic with $10^{-3}$ and $10^{-4}$ [Gy/hr] dose-rate groups, respectively. The calculated results are indicated by three lines, straight and dotted and dashed -dotted lines corresponding to acute ($5 \times 10$ [Gy/hr]), chronic with $5 \times 10^{-3}$ [Gy/hr]), chronic with ($5 \times 10^{-4}$ [Gy/hr]) groups, respectively.

Fig. 8. The comparison between experimental data points with error bars and theoretical values of mutation frequency.

Table 1.   Specific locus mutation rate data obtained in $A_s$ spermatogonia by use of the seven-locus test stock provided by ref. 4

| Label | Category | Source of radiation | Total dose (Gy) | dose-rate (Gy/hour) | Mutations, no. | Offspring, no. | Mutaion Frequency $\times 10^5$ per locus | Ref. |
|---|---|---|---|---|---|---|---|---|
|  | Control |  |  |  | 28 | 531,500 | 0.75 | 7 |
|  |  |  |  |  | 11 | 157,421 | 1.00 | 8 |
|  |  |  |  |  | 0 | 38,448 | 0.00 | 9 |
| A | Chronic | $^{137}$Cs | 3.00 | 0.00042 | 11 | 48,358 | 3.25 | 3 |
| B |  | $^{60}$Co | 0.38 | 0.00060 | 7 | 79,364 | 1.26 | 8 |
| C |  | $^{137}$Cs | 0.86 | 0.00060 | 6 | 59,810 | 1.43 | 7 |
| D |  | $^{137}$Cs | 3.00 | 0.00060 | 15 | 49,569 | 4.32 | 7 |
| E |  | $^{137}$Cs | 6.00 | 0.00060 | 22 | 53,380 | 5.89 | 7 |
| F |  | $^{137}$Cs | 3.00 | 0.0030 | 24 | 84,831 | 4.04 | 3 |
| G |  | $^{60}$Co | 6.71 | 0.0030 | 20 | 58,795 | 4.86 | 8 |
| H |  | $^{60}$Co | 6.18 | 0.0048 | 5 | 22,682 | 3.15 | 8 |
| I |  | $^{137}$Cs | 3.00 | 0.0054 | 10 | 58,457 | 2.44 | 7 |
| J |  | $^{137}$Cs | 5.16 | 0.0054 | 5 | 26,325 | 2.71 | 7 |
| K |  | $^{137}$Cs | 8.61 | 0.0054 | 12 | 24,281 | 7.06 | 7 |
| L |  | $^{137}$Cs | 6.00 | 0.480 | 10 | 28,059 | 5.09 | 10 |
| M | Acute | X-ray | 3.00 | 54.000 | 40 | 65,548 | 8.72 | 7 |
| N |  | X-ray | 6.00 | 54.000 | 111 | 119,326 | 13.29 | 7 |
| O |  | X-ray | 6.70 | 43.200 | 12 | 11,138 | 15.39 | 8 |

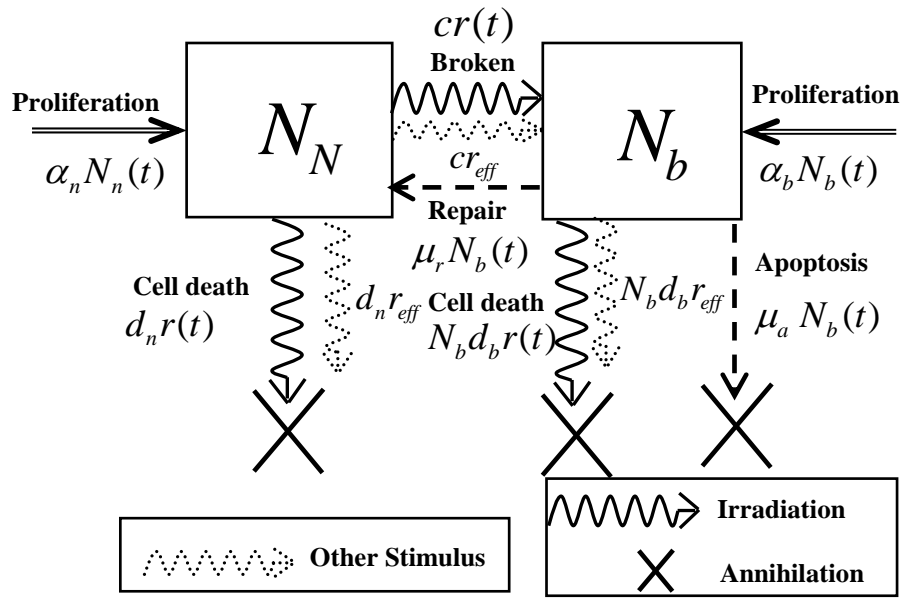

Fig. 1. Stimulus response diagram of LD model

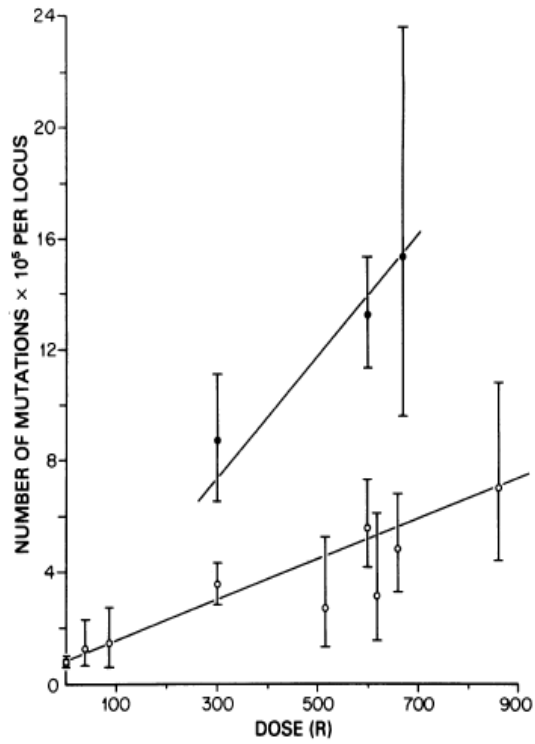

Fig. 2. Mutation rates (multiplied by $10^5$) at seven specific loci in the mouse versus total dose. Original figure presented by ref. 4 is to be compared with the one with all the individual data points listed in Table 1 and the unit of dose is expressed in terms of roentgen unit. The straight lines are obtained by ref. 4, the best estimate of the ratio of mutation frequencies for chronic and acute irradiation.

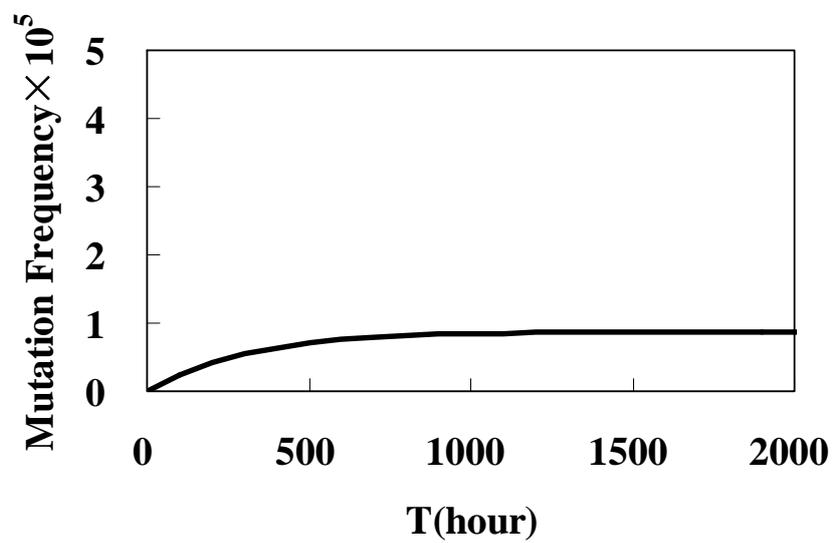

Fig. 3. An example of time dependence of Eq. (4.2).

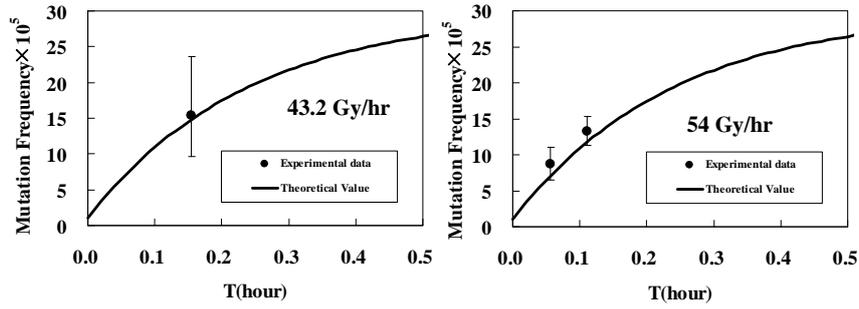

Fig. 4-a

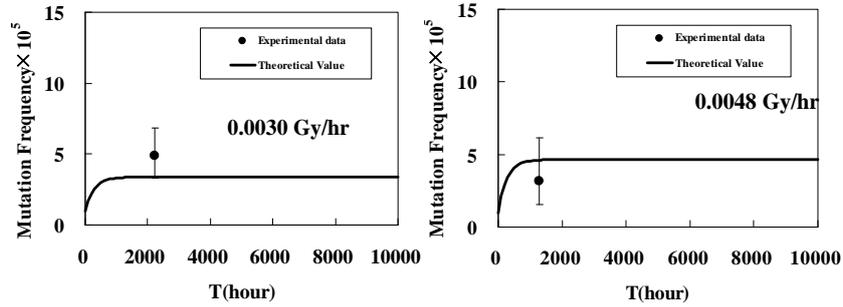

Fig. 4-b

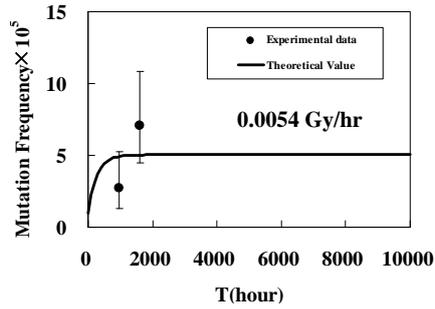

Fig. 4-c

Fig.4. Time dependence of mutation frequency dividing into each dose rate. The data points are indicated by those predicted at time T. T is determined by $T = (\text{total dose}) / (\text{dose rate})$ which corresponds the time when it reaches the corresponding total dose of each data.

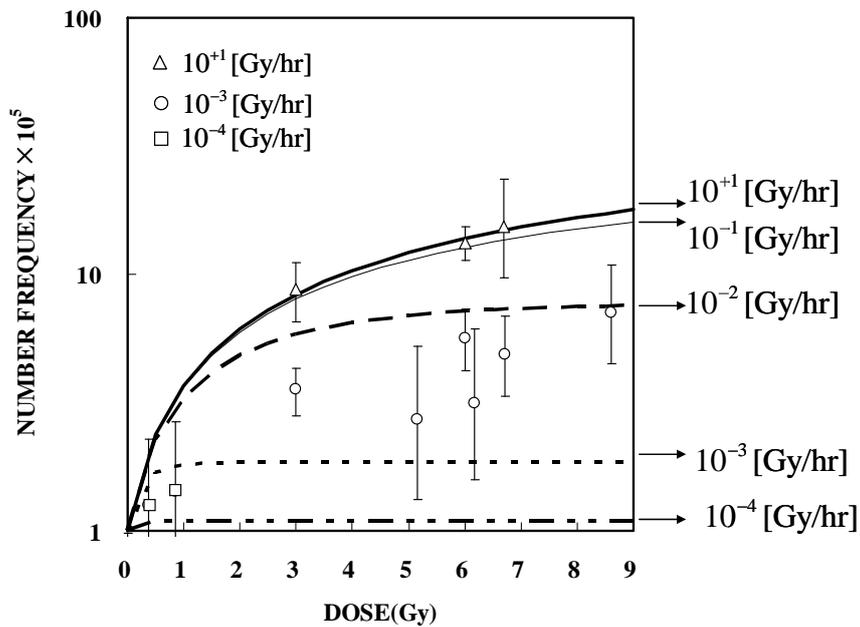

Fig.5. Explicit dose rate dependence of mutation frequency $F$ can be easily seen by dividing the whole data points into groups with the same order dose rate groups, acute, chronic with $10^{-3}$ and $10^{-4}$ [Gy/hr] dose-rate groups, respectively. The calculated results are indicated by three lines, straight and dotted and dashed -dotted lines corresponding to acute ($5 \times 10$ [Gy/hr]), chronic with $5 \times 10^{-3}$ [Gy/hr]), chronic with ($5 \times 10^{-4}$ [Gy/hr]) groups, respectively. Nine data points are also indicated in the figure with error bars.

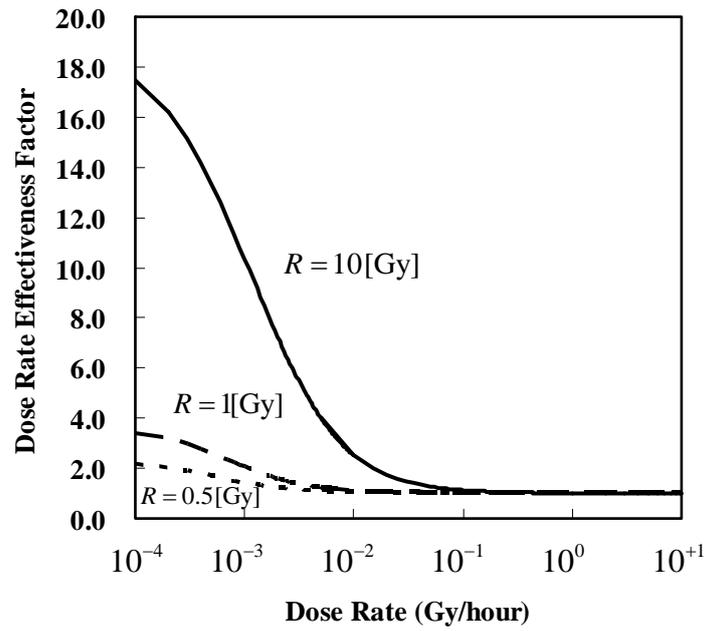

Fig. 6. Dose rate dependence of DREF. Solid line, dashaed line and dotted line correspond to the case of total dose $R = 10, 1, 0.5 [\text{Gy}]$, respectively.

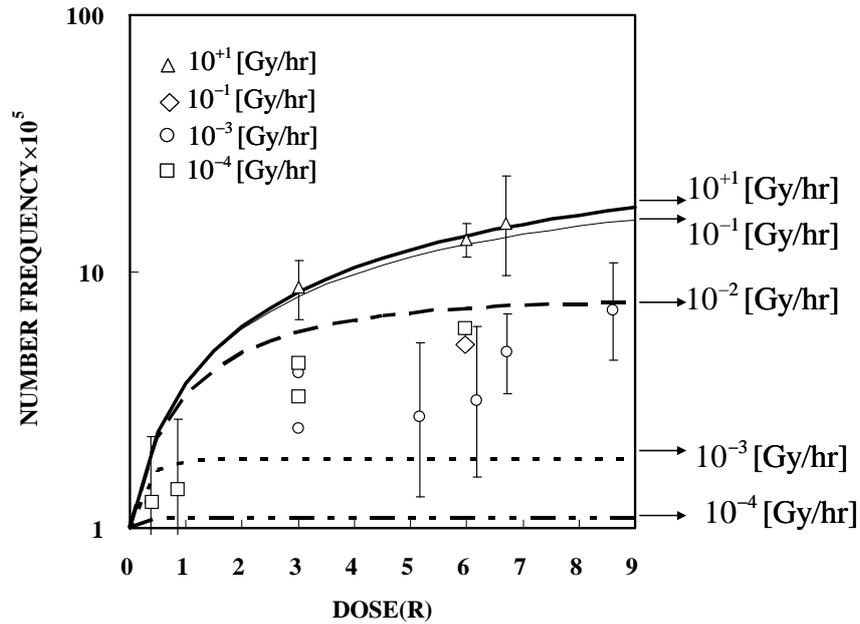

Fig.7. Explicit dose rate dependence of mutation frequency  *F* can be easily seen by dividing the whole data points into groups with the same order dose rate groups, acute, chronic with  $10^{-3}$  and  $10^{-4}$  [Gy/hr] dose-rate groups, respectively. The calculated results are indicated by three lines, straight and dotted and dashed -dotted lines corresponding to acute ($5\times 10$ [Gy/hr] ), chronic   with  $5\times 10^{-3}$ [Gy/hr] ), chronic with ($5\times 10^{-4}$ [Gy/hr] ) groups, respectively.

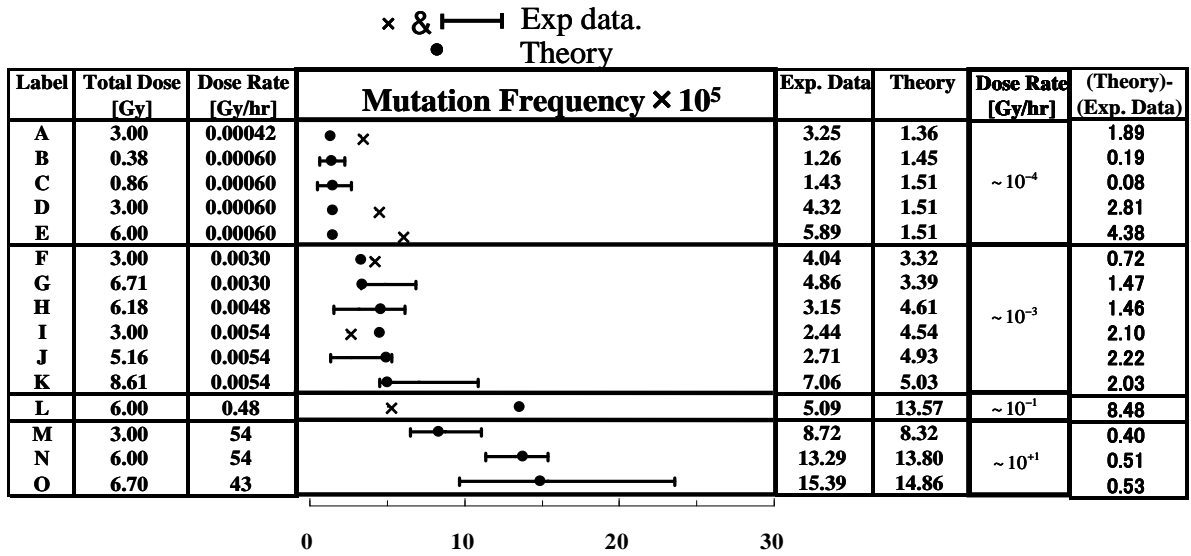

| Label | Total Dose [Gy] | Dose Rate [Gy/hr] | Mutation Frequency × 10⁵ | Exp. Data | Theory | Dose Rate [Gy/hr] | (Theory)-(Exp. Data) |
|---|---|---|---|---|---|---|---|
| A | 3.00 | 0.00042 | | 3.25 | 1.36 | | 1.89 |
| B | 0.38 | 0.00060 | | 1.26 | 1.45 | | 0.19 |
| C | 0.86 | 0.00060 | | 1.43 | 1.51 | $\sim 10^{-4}$ | 0.08 |
| D | 3.00 | 0.00060 | | 4.32 | 1.51 | | 2.81 |
| E | 6.00 | 0.00060 | | 5.89 | 1.51 | | 4.38 |
| F | 3.00 | 0.0030 | | 4.04 | 3.32 | | 0.72 |
| G | 6.71 | 0.0030 | | 4.86 | 3.39 | | 1.47 |
| H | 6.18 | 0.0048 | | 3.15 | 4.61 | $\sim 10^{-3}$ | 1.46 |
| I | 3.00 | 0.0054 | | 2.44 | 4.54 | | 2.10 |
| J | 5.16 | 0.0054 | | 2.71 | 4.93 | | 2.22 |
| K | 8.61 | 0.0054 | | 7.06 | 5.03 | | 2.03 |
| L | 6.00 | 0.48 | | 5.09 | 13.57 | $\sim 10^{-1}$ | 8.48 |
| M | 3.00 | 54 | | 8.72 | 8.32 | | 0.40 |
| N | 6.00 | 54 | | 13.29 | 13.80 | $\sim 10^{+1}$ | 0.51 |
| O | 6.70 | 43 | | 15.39 | 14.86 | | 0.53 |

Fig. 8. The comparison between experimental data points with error bars and theoretical values of mutation frequency.